\documentclass[12pt]{article}
\usepackage{mathptmx}
\usepackage{iopconf_lorente}
\usepackage{graphicx}
\usepackage{bm}
\usepackage{amsmath}
\usepackage{amssymb}

\begin{document}
\title{Discrete quantum gravity: The Lorentz invariant weight for the Barrett-Crane model}

\author{Miguel Lorente}

\affil{\ Departamento de F\'{\i}sica, Univ. de Oviedo, 33007 Oviedo, Spain}

\affil{\ Inst. f\"{u}r theor. Physik, Univ. Tuebingen, 72076 Tuebingen, Germany}

\beginabstract 
In a recent paper [1] we have constructed the spin and tensor representations of $SO(4)$ from which
the invariant weight can be derived for the Barrett-Crane model in quantum gravity. By analogy with
the $SO(4)$ group, we present the complexified Clebsch-Gordan coefficients in order to construct the
Biedenharn-Dolginov function for the $SO(3,1)$ group and the spherical function as the Lorentz
invariant weight of the model.
\endabstract

\section{Review of the Euclidian model}
Given a triangulation of a 4-dimensional Riemannian manifold, we assigne bivectors to the faces
satisfying appropiate constraints [2]. Then we identify the bivectors with Lie algebra elements and
associate a representation of $SO(4)$ to each triangle and a tensor to each tetrahedrum, invariant
under $SO(4)$. The representation chosen is simple, i.e. $j_1=j_2$. 

Now it is easy to construct an amplitude for the quantum 4-simplex. The graph for a spin foam is the
1-complex, dual to the boundary of the 4-simplex having five 4-valent vertices (corresponding to the
five tetrahedra) with each of the ten edges connecting two different vertices (corresponding to the
ten triangles of the 4-simplex each shared by two tetrahedra). Now we associate to each triangle (the
dual of which is an edge) a simple representation of the algebra of $SO(4)$ and to each tetrahedra
(the dual of which is a vertex) an intertwiner; and to a 4-simplex the product of the five
intertwiners and the sum for all possible representations. The proposed state sum is:
$$Z=\sum\limits_J {\prod\limits_{triang.} {A_{tr}}}\prod\limits_{tetrahedra} {A_{tetr.}}\prod\limits_{4-simplex} {A_{simp.}}$$
For the simple representations $(j_1=j_2)$ attached to every face of the tetrahedrum, we used the elementary spherical
functions, that can be calculated from the Biedenharn-Dolginov function in the case $j=m=0$, namely, 
$$d_{000}^{(j_1,0)}(\tau )=\frac{{{\sin (2j_1+1)\tau}}}{{(2j_1+1)\sin \tau }}$$

\section{Spinor representation of SL(2,C)}

We define the complex valued polynomials
$$p(z,\bar z)=\sum {C_{\alpha \beta }z^\alpha }\bar z^\beta $$
as the basic states of the spinor representations:
$$T_ap(z,\bar z)=(a_{10}z+a_{11})^k(\bar a_{10}\bar z+\bar a_{11})^np\left( \frac{{{a_{00}z+a_{01}}}} {{a_{10}z+a_{11}}},\frac{{{\bar a_{00}\bar
z+\bar a_{01}}}}{\bar a_{10}\bar z+\bar a_{11}} \right)$$
for any $a\in SL(2,\mathbb{C},\;\;a\equiv \left( \begin{matrix}a_{00}& a_{01}\\ a_{10}& a_{11}\end{matrix} \right) .$

This representation with labels $\left( {l_0,l_1} \right)=\left( \frac{k-n}{2},\frac{k+n}{2}+1 \right),\ k,\;n\in
\mathbb{N}$, is irreducible, and finite dimensional. If we enlarge this representation with complex values, $k,n$, we get
$$T_af\left( {z,\bar z} \right)=\left( {a_{10}z+a_{11}} \right)^{l_0+l_1-1}\left( {\bar a_{10}\bar z+\bar a_{11}}
\right)^{l_1-l_0-1}f\left( {{{a_{00}z+a_{01}} \over {a_{10}z+a_{11}}},{{\bar a_{00}\bar z+\bar a_{01}} \over {\bar a_{10}\bar z+\bar a_{11}}}}
\right),$$ 
the representation becomes infinite dimensional. For $l_0$ integer or half integer and $l_1$ complex the representation is irreducible. The
principal series of unitary representations for $SL(2,\mathbb{C})$ are defined, on the Hilbert space of complex functions, with scalar
product $\left( {f_1,f_2} \right)=\int {f_1}(z)f_2(z)\;dz$, as $$T_af\left( z \right)=\left( {a_{10}z+a_{11}} \right)^{\mu -1+i\gamma }\left(
{\bar a_{10}\bar z+\bar a_{11}} \right)^{-\mu -1+i\gamma }f\left( {{{a_{00}z+a_{01}} \over {a_{10}z+a_{11}}}} \right)$$
with $l_0= \mu$, integer or half integer, $l_1= i\gamma$, $\gamma \in \mathbb{R}$.

The complementary series of unitary representations are defined in the Hilbert space of complex functions, with scalar product $\left(
{f_1,f_2} \right)=\int {\left| {z_1-z_2} \right|}^{-2+2\sigma }f_1\left( {z_1} \right)f_2\left( {z_2} \right)\;dz_1dz_2,$ as
$$T_af\left( z \right)=\left| {a_{10}z+a_{11}} \right|^{2\sigma -2}f\left( {{{a_{00}z+a_{01}} \over {a_{10}z+a_{11}}}} \right)$$
with $l_0=0,\;\;l_1=\sigma ,\;\;\sigma \in \mathbb{R},\;\;\left| \sigma  \right|<1.$

\section{Representations of the algebra of SL(2,$\mathbb{C}$)}
Given the generators of rotations $\left( {J_1,J_2,J_3} \right)=\bar J$ and of pure Lorentz transformations $\left(
{K_1,K_2,K_3} \right)=\bar K$ satisfying the commutation relations:
$$\begin{array}{llllll}
&\left[ {J_p,J_q} \right]=i\varepsilon_{pqr}\,J_r\quad &,\quad &\bar J^+=\bar J\quad &,\quad &p,q,r=1,2,3\cr
  &\left[ {J_p,K_q} \right]=i\varepsilon_{pqr}\,K_r\quad &,\quad &\bar K^+=\bar K\cr
  &\left[ {K_p,K_q} \right]=-i\varepsilon_{pqr}\,J_r\cr
\end{array}$$
we obtain the unitary representations of the algebra of $SL(2, \mathbb{C})$ in the basis where the operators $J_3$ and $\bar J^2$ are diagonal,
namely: $J_3\psi _{jm}=m\psi _{jm},\quad \bar J^2\psi _{jm}=j(j+1)\psi _{jm}$

It is possible also to construct a complexified operators
$$\bar A={1 \over 2}\left( {\bar J+i\bar K} \right),\quad \bar B={1 \over 2}\left( {\bar J-i\bar K} \right),\quad \bar A^+=\bar B$$
that leads to the commutation relations of two independent angular momenta:

\vspace{0,1 cm}
$\left[ {A_p,A_q} \right]=i\varepsilon_{pqr}\,A_r$

$\left[ {B_p,B_q} \right]=i\varepsilon_{pqr}\,B_r$

$\left[ {A_p,B_q} \right]=0$
\vspace{0,2 cm}

Since $J_3$ and $K_3$ commute we construct the representations of these operators in the basis where $J_3$ and $K_3$ are diagonal, [3]

\vspace{0,1 cm}
$J_3\phi _{m_1m_2}=m\phi _{m_1m_2}\quad ,\quad K_3\phi _{m_1m_2}=\lambda \phi _{m_1m_2},$ hence

$A_3\phi _{m_1m_2}={1 \over 2}\left( {m+i\lambda } \right)\phi _{m_1m_2}\equiv m_1\phi _{m_1m_2}$

$B_3\phi _{m_1m_2}={1 \over 2}\left( {m-i\lambda } \right)\phi _{m_1m_2}\equiv m_2\phi _{m_1m_2}$
\vspace{0,2 cm}

Notice that $\lambda$ is a real continuous parameter, but $m_1$ and $m_2$ are complex conjugate and $\bar m_1=m_2$

In both basis the labels of the representations $(l_0,l_1)$ takes the values $l_0=\mu $ (integer or half integer),
$l_1=i\gamma (\gamma \in R)$ for the principal series, and $l_0=0,\; l_1=\sigma ,\; \left| \sigma  \right|<1,\;\sigma \in R$, for
the complementary series. For the Casimir operators we have

\vspace{0,1 cm}
$\left( {\bar J^2- K^2} \right)\psi _{jm}=\left( {l_0^2+l_1^2-1} \right)\psi _{jm}$

$\left( {\bar J\cdot \bar K} \right)\psi _{jm}=l_0l_1\psi _{jm}$

\section{Complexified Clebsch-Gordan coefficients and the representation of the boost operator}

In order to connect the basis $\psi _{jm}$ and $\phi _{m_1m_2}$ we can use the complexified Clebsch-Gordan coefficients:
\begin{equation}
\psi _{jm}=\int\limits_{-\infty }^\infty  {d\lambda \left\langle {{m_1m_2}} \mathrel{\left | {\vphantom {{m_1m_2} {j_m}}} \right.
\kern-\nulldelimiterspace} {{jm}} \right\rangle }\phi _{m_1m_2}
\end{equation}
We have used integration because $\lambda$  is a continuous parameter. It can be proved that these coefficientes are related to the Hahn
polynomials of imaginary argument [3]
\begin{equation}
\left\langle {{m_1m_2}} \mathrel{\left | {\vphantom {{m_1m_2} {j_m}}} \right. \kern-\nulldelimiterspace} {{jm}} \right\rangle =f{{\sqrt {\rho
(\lambda )}} \over {d_{j-m}}}p_{j-m}^{(m-\mu ,m+\mu )}(\lambda ,\gamma )\quad , \quad f \bar f=1,
\end{equation}
\hspace{300pt}for the principal series,
\begin{equation}
\left\langle {{m_1m_2}} \mathrel{\left | {\vphantom {{m_1m_2} {j_m}}} \right. \kern-\nulldelimiterspace} {{jm}} \right\rangle =f\sqrt {\rho
(\lambda )}\;d_{j-m}^{-1}\;q_{j-m}^{(m)}(\lambda ,\sigma )
\end{equation}
\hspace{300pt}for the complementary series,

\noindent where
\begin{equation}
\rho (\lambda )={1 \over {4\pi }}\left| {\Gamma \left( {{{m-\mu +1} \over 2}+i{{\lambda -\gamma } \over 2}} \right)\Gamma \left( {{{m+\mu +1}
\over 2}+i{{\lambda +\gamma } \over 2}} \right)} \right|^2
\end{equation}
\noindent and

\begin{equation*}
d_n^2={{\Gamma \left( {m-\mu +n+1} \right)\Gamma \left( {m+\mu +n+1} \right)\left| {\Gamma \left( {m+i\gamma +n+1}
\right)} \right|^2} \over {n!\;\left( {2m +2n+1} \right)\Gamma \left( {2m +n+1} \right)}}
\end{equation*}
for the principal series and similar expression for the complementary series. 

With the help of equations (1), (2) and (3) we can construct the
representation for the boost operator, or the Biedenharn-Dolginov function, namely,
\begin{align*}
d_{jj'm}^{(\mu ,\gamma )}(\tau )&=\left\langle {{\psi _{jm}}} \mathrel{\left | {\vphantom {{\psi _{jm}}
{e^{-i\tau K_3}\psi _{j'm}}}} \right.
\kern-\nulldelimiterspace} {{e^{-i\tau K_3}\psi _{j'm}}} \right\rangle =\\
&=\int\limits_{-\infty }^\infty 
{d_{j-m}^{-1}}p_{j-m}^{(m-\mu ,m+\mu )}(\lambda ,\gamma )e^{-i\tau \lambda }d_{j'-m}^{-1}p_{j'-m}^{(m-\mu ,m+\mu )}(\lambda ,\gamma
)\rho (\lambda )d\lambda
\end{align*}

\section{Spherical function for the group SO(3,1)}

Given a locally compact group $G$ with completely irrep. $T_g$, and a compact subgroup $K\subset G$, with completely irrep.
$T_k$ (finite dimensional), we define the spherical function 
$$f(g)=T_r\left\{ {E_kT_g} \right\}=T_r\left\{ {T_k} \right\}$$
\noindent where $E_k$ is the projector $E_k:T_g\to T_k.$

The spherical fucntions [4] are functions on the homogeneous space $K \verb"\G"$ and invariant on right cosets:
$f(kg)=f(g)$

We apply this definition to the representation of $SO(3,1)$ given by Biedenharn function, with $\mu = 0$, projected into the
identity representation of $SO(3)$, with $j=m=0$ (the elementary spherical function). We have 

\begin{align*}
f(\tau )&=Trd_{000}^{0,\gamma }=\int {d_0^{-2}}e^{-i\lambda \tau }\left[ {p_0^{(0,0)}} \right]^2\rho (\lambda )d\lambda
=\\ 
&=\int\limits_{-\infty }^\infty  {{{e^{-i\lambda \tau }} \over {\left| {\Gamma (1+i\gamma )} \right|^2}}}{1 \over {4\pi
}}\left| {\Gamma \left( {{1 \over 2}+i{{\lambda +\gamma } \over 2}} \right)\Gamma \left( {{1 \over 2}+i{{\lambda -\gamma }
\over 2}} \right)} \right|^2d\lambda =\\
&=\int\limits_{-\infty }^\infty  {e^{-i\lambda \tau }}{{sh\pi \gamma } \over {4\gamma }}{1
\over {ch\pi \left( {{{\lambda +\gamma } \over 2}} \right)}}{1 \over {ch\pi \left( {{{\lambda -\gamma } \over 2}}
\right)}}d\lambda
\end{align*}

\noindent where we have used the properties of ${\Gamma }$ functions. From the residue theorem at the poles $\lambda
=\mp \gamma +\left( {2n+1} \right)i,\;\;n=0,1,2,\dots,$ we easily obtain

$$f(\tau )=i{{e^{i\gamma \tau }-e^{-i\gamma \tau }} \over \gamma }e^\tau \sum\limits_{n=0}^\infty  {\left( {e^{2\tau }} \right)^n}={1 \over \gamma }{{\sin
\gamma \tau } \over {sh\tau }}$$ for the principal series with $l_0=0,\;l_1=i\gamma .$ We obtain by the same way
$$f(\tau )={1 \over \sigma }{{sh\sigma \tau } \over {sh\tau }}$$
for the complementary series, with $l_0=0,\;l_1=\sigma ,\;\left| \sigma  \right|<1$

\section{A SO(3,1) invariant for the state sum of spin foam model} 

As in the case of euclidean $SO(4)$ invariant model, we take a non degenerate finite triangulation of a 4-manifold. We consider
the 4-simplices in the homogeneous space $SO(3,1)/SO(3)$ $\backsim H_3$, the hyperboloid $\left\{ {\left. x \right|x\cdot
x=1,x^0>0} \right\}$ and define the bivectors $b$ on each face of the 4-simplex, that can be space-like, null or timelike
$\left( {\left\langle {b,b} \right\rangle >0,=0,<0,\makebox {respectively}} \right)$.

In order to quantize the bivectors, we take the isomorphism $b=*L\left( {b^{ab}={1 \over
2}\varepsilon^{abcd}L_d^eg_{ec}} \right)$ with $g$ a Minkowski metric.

The condition for $b$ to be a simple bivector $\left\langle {b,*b} \right\rangle =0$, gives $C_2=\left\langle {L,*L}
\right\rangle =\bar J\cdot \bar K=\mu \gamma =0$

We have two cases:

1) $\gamma =0,\quad C_1=\left\langle {L,L} \right\rangle =\mu ^2-1>0$; $L$, space time, $b$ time-like

2) $\mu =0,\quad C_1=J^2-K^2=-\gamma ^2-1<0$; $L$, time like, $b$ space like (remember, the Hodge operator $*$ changes the
signature)

In case 2) $b$ is space-like, $\left\langle {b,b} \right\rangle >0$. Expanding this expression in terms of space like vectors,
$x,y,$ 
\begin{align*}
b_{\mu \nu }b^{\mu \nu }&=\left( {x_\mu y_\nu -x_\nu y_\mu } \right)\left( {x^\mu y^\nu -x^\nu y^\mu } \right)=\\
&=\left\| x
\right\|^2\left\| y \right\|^2-\left\| x \right\|^2\left\| y \right\|^2\cos ^2\eta \left( {x,y} \right)=\left\| x
\right\|^2\left\| y \right\|^2\sin ^2\eta \left( {x,y} \right)
\end{align*}
where $\eta \left( {x,y} \right)$ is the Lorentzian space-like angle between $x$ and $y$; this result gives a geometric
interpretation between the parameter $\gamma$ and the area expanded by the bivector $b=x\wedge y$, namely, $\left\langle {b,b}
\right\rangle =\mbox {area}^2\left\{ {x,y} \right\}=\left\langle {*L,*L} \right\rangle \cong \gamma ^2$. (This result is
equivalent to that obtained in [1] where the area of the triangle expanded by the bivector was proportional to the value $(2j+1)$,
$j$ being the spin of the representation).

In order to have the state sum we attach to each face of the 4-simplex the spherical function described in Section 5 and
integrate over the whole rank of the parameter of the representation. The resulting partition function is invariant under
$SO(3,1)$ and has been proved to be finite [5]. The assymptotic properties of the spherical function are related to the
Einstein-Hilbert action, giving a connection of the model with the theory of general relativity.

\section*{Acknowledgments}
The author expresses his gratitude to the Director of the Institut f\"{u}r theor. Physik, of the University of Tuebingen, where part of this
work was done, and to Prof. Barrett for iluminating conversation about spherical functions. This work has been partially supported by M.E.C. (Spain).
Grant: BFM2003--00313/FIS.

\end{document}